\documentstyle[prd,aps,eqsecnum]{revtex}
\begin{document}
\title{\bf Form-factors  of the sausage model obtained\\
with bootstrap fusion from sine-Gordon theory}
\author{ Zal\'an Horv\'ath \\
         Institute for Theoretical Physics\\
         Roland E\"otv\"os University, Budapest, Hungary\\
         G\'abor Tak\'acs \\
         Department of Applied Mathematics and Theoretical Physics\\
         Cambridge, UK }
\date{6 November 1995}
\maketitle
\begin{abstract}
We continue the investigation of massive integrable models by  
means of the bootstrap fusion procedure, started in our previous 
work on O(3) nonlinear sigma model. 
Using the analogy with SU(2) Thirring model 
and the O(3) nonlinear sigma model we prove a similar relation 
between sine-Gordon theory and a one-parameter deformation of 
the O(3) sigma model, the sausage model. This allows us to write 
down a free field representation for the Zamolodchikov-Faddeev 
algebra of the sausage model and to construct an integral 
representation for the generating functions of form-factors 
in this theory. We also clear up the origin of the singularities 
in the bootstrap construction and the reason for the problem 
with the kinematical poles.
\end{abstract}
\section{Introduction}\label{sect1}

Exactly solvable two-dimensional quantum field theories have 
attracted a great deal of interest and research work recently. 
There are several reasons for that: these theories can serve 
as laboratory for higher dimensional theories; there are 
physical systems with effectively two-dimensional phenomena (eg. 
quantum Hall effect); and there is the huge and rapidly developing 
area of string theory, topological field theory and related topics, 
in which two-dimensional physics plays a central role.

A large variety of different approaches emerged in trying to solve 
these models of quantum field theory. The first real breakthrough 
was achieved in \cite{zamzam}, in which exact S-matrices were 
conjectured for several interesting models using bootstrap ideas.
The fact that made this possible is that scattering theory of 
quantum integrable models is very simple: it is factorized 
scattering theory (FST). This means that the many-particle 
S-matrix can be written in terms of products of two-particle 
S-matrices. The consistency condition for this turns out to 
be the famous Yang-Baxter equation, which, together with unitarity,
crossing symmetry and the spectrum of the particles of the model, 
severely restricts the class of possible S-matrices that one 
can write down. However, there is a so-called CDD ambiguity, which means 
that the S-matrix is not determined uniquely by the aforementioned 
conditions. In most cases one solves this difficulty by choosing 
a 'minimal' solution to the factorization equations. This choice 
can be justified in many cases by comparison of results coming 
from the conjectured S-matrix and from the Lagrangian field theory.

The S-matrix, however, describes only the on-shell 
physics. For off-shell physics, one turns to other quantities  
which contain more information: correlation functions. For their 
calculation the knowledge of the matrix elements of local 
operator is needed. These matrix elements are called 
form-factors; correlation functions can be represented as an 
infinite series constructed using these objects. For the 
majority of integrable field theories this is the only tool 
to compute correlation functions up to now; in some simple cases
one can use other techniques. 

Form-factors can be found solving a very complicated system 
of axioms, which have been known for a long time \cite{berg}. 
The two-particle minimal form-factors were calculated for 
many models, however the calculation of the many-particle 
ones posed essential difficulties. Then, in a remarkable series 
of papers, Smirnov \cite{smirnov1,smirnov2,smirnov3,smirnov4} 
obtained all form-factors of some important operators in 
the sine-Gordon, SU(2) Thirring and O(3) sigma models. 
Further investigation revealed that the key structure behind 
these solutions is that of quantum affine algebras, which is 
a nonlocal symmetry of these integrable field theories 
\cite{bernard}. The Zamolodchikov-Faddeev operators (which 
play a central role in our approach as well) turned out to 
be vertex operators for the representations of the nonlocal 
symmetry algebra \cite{smirnov5}.

The analogy between the Knizhnik-Zamolodchikov (KZ) equations 
for affine Kac-Moody algebras and the form-factor axioms, 
one of which proved to be a quantum analogue of the KZ 
equation \cite{smirnov5,frenkel}, as well as the vertex 
operator representation of quantum affine algebras \cite{jing}
strongly suggest that a bosonization technique based on free 
fields, similar to the one used in conformal field theory, 
should be useful in solving the form-factor equations. This 
is the central idea in the papers of Lukyanov 
\cite{freeaffin,freefield}, where he was able to construct a free 
field representation for the sine-Gordon and SU(2) Thirring 
model and to obtain an integral representation for the form 
factors. In our earlier work \cite{prev}, using mainly his 
results and the connection between SU(2) Thirring and O(3) 
sigma models put forward by Smirnov, we defined a free field 
representation for the O(3) model and applied it to build up an 
integral representation for the form-factors. Here we extend 
the method to the sine-Gordon and sausage models and in this 
way write down a similar representation for the form-factors 
of the sausage model. The problems encountered in \cite{prev} 
when applying the method unfortunately remain the same and 
we choose the same old solution for correcting the analytic 
structure introducing some multiplier functions. Our main 
result is to obtain a procedure following which one can 
calculate form-factors of the sausage model. We also provide 
a detailed analysis of the analytical structure and an 
interpretation of the double poles in terms of an analogy 
with thermal Green functions. The relation of the 
problem of the kinematical poles and the singularity of 
the fusion limit of the sine-Gordon form-factor is cleared 
up.

The paper is organized as follows. Section II reviews 
factorized scattering theory of sine-Gordon and sausage 
models and there we discuss the quantum symmetries. In Section 
III we briefly describe the form-factor axioms and the free 
field representation for their solution. Section IV 
deals with the question of the bootstrap fusion procedure used 
for transition from the sine-Gordon model to the sausage model
and using these results  defines the free field 
representation for the sausage model, while in Section V we 
give the general formulas for evaluating the form-factors 
and provide a detailed discussion  of their analytic structure. 
Section VI is reserved for the conclusions.

We added three appendices to the paper. 
Appendix A describes the method used 
to obtain the bootstrap fusion and the Zamolodchikov-Faddeev 
algebra relation on the level of the free field representation, 
while Appendix B shows how to evaulate the traces occurring 
in the calculation of generating functions with coherent 
oscillator methods. In Appendix C we describe a sample calculation 
of the operator product singularities.

\section{Review of the sine-Gordon and sausage field theories}
\label{sect2} 

Sine-Gordon theory is defined by the Lagrangian density
\begin{equation}
{\cal L}={1\over 2}(\partial_\mu\phi )^2+{m_0^2\over b^2}\cos (b\phi )
\ ,\label{sglagr}\end{equation}
where $b$ is a coupling constant and $m_0$ is a mass parameter.
Instead of $b$ we will use the parameter
\begin{equation}
\xi={b^2\over 8\pi-b^2}\ ,
\label{xidef}\end{equation}
which can be taken as the only dimensionless parameter of the 
model. When $\xi >1$, the quantum spectrum of the theory contains 
only a doublet of solitons which forms the fundamental representation 
of the quantum group $SU(2)_q$ with
\begin{equation}
q=\exp [i\pi(1+{1\over\xi})]\ .
\label{qxi}\end{equation}
When $\xi <1$, the spectrum contains breathers, which are bound states 
of the fundamental solitons. We do not treat this case here. 
We also restrict ourselves to study the case when $q$ is not a root of 
unity to avoid dealing with degenerate representations of the 
quantum group. The S-matrix 
of the solitons was found in \cite{zamzam} and looks as follows:
\begin{eqnarray} 
&&S^{++}_{++}(\beta ) = S^{--}_{--}(\beta )=S_0(\beta ), \nonumber \\
&&S^{+-}_{+-}(\beta ) = S^{-+}_{-+}(\beta )=S_0(\beta )
{\sinh({\beta\over\xi})\over \sinh({i\pi -\beta\over\xi})}, \nonumber \\
&&S^{+-}_{-+}(\beta ) = S^{-+}_{+-}(\beta )=S_0(\beta )
{\sinh({i\pi\over\xi})\over \sinh({i\pi -\beta\over\xi})},
\label{sgsmatrix}\end{eqnarray}
where $+,-$ is the internal index, 
$\beta$ is the rapidity parameter, which is
related to the energy-momentum by $p_0=m\cosh\beta,\ p_1=m\sinh\beta$, 
m is the soliton mass and 
\begin{equation}
S_0(\beta )=-\exp\bigg[ -i\int\limits_0^\infty{\sin (\kappa\beta )\sinh 
({1-\xi\over2}\pi\kappa )d\kappa\over \kappa\cosh({1\over 2}\pi\kappa)
\sinh({\xi\over 2}\pi\kappa)}\bigg].
\label{s0def}\end{equation}

The Hilbert space of the theory can be described by introducing the 
formal Zamolodchikov-Faddeev operators $Z_i(\beta )$ satisfying the 
following algebra relations 
\begin{equation}
Z_i(\beta_1)Z_j(\beta_2)= S_{ij}^{kl}(\beta_{12})Z_l(\beta_2)
Z_k(\beta_1), \qquad \beta_{12}=\beta_1-\beta_2.
\label{scattrel}\end{equation}
As we shall see below, the commutation relations of the Z-operators 
reflect the scattering of the particles. From now on we will refer 
to (\ref{scattrel}) as the scattering relations for the sine-Gordon 
model. The Hilbert space furnishes a representation of the 
Zamolodchikov-Faddeev (ZF) algebra. The ZF operators create the 
asymptotic in and out states:
\begin{eqnarray}
\vert Z_{i_1}(\beta_1)...Z_{i_n}(\beta_n)\rangle_{out}=
Z_{i_1}(\beta_1)...Z_{i_n}(\beta_n)\vert{\rm vac}\rangle , \nonumber \\
\vert Z_{i_1}(\beta_1)...Z_{i_n}(\beta_n)\rangle_{in}=
Z_{i_n}(\beta_n)...Z_{i_1}(\beta_1)\vert{\rm vac}\rangle\ ,
\label{definout}\end{eqnarray}
where $\vert{\rm vac}\rangle$ is the vacuum state and the rapidities 
are ordered as $\beta_1<\beta_2<...<\beta_n$. Now if one expresses the 
out states in terms of the in states, one gets the $n$-particle 
S-matrix as a product of the 2-particle S-matrices. The associativity 
of the ZF algebra then implies the famous Yang-Baxter equation, which 
is the factorization condition for the S-matrix \cite{zamzam}. 

The Hilbert space structure is specified by defining the conjugate ZF 
operators $Z^{i\dagger}$, which annihilate the vacuum and satisfy the 
following additional relations
\begin{eqnarray}
&&Z^{i\dagger}(\beta_1)Z^{j\dagger}(\beta_2) =
Z^{l\dagger}(\beta_2)Z^{k\dagger}(\beta_1)
S^{ij}_{kl}(\beta_{12}) \ ,\nonumber \\
&&Z_i(\beta_1)Z^{j\dagger}(\beta_2) =
Z^{k\dagger}(\beta_2)S^{jl}_{ik}
(\beta_{12})Z_l(\beta_1)+2\pi\delta^j_i\delta(\beta_{12}).
\label{zdagrel}\end{eqnarray}
However, in the sequel we will deal with another representation of the 
ZF algebra defined by (\ref{scattrel}) which does not include the 
conjugate ZF operators and therefore it is different from the physical 
Hilbert space of the model. This representation will be the space of 
the free field system used to solve the form-factor equations. 

As mentioned above, the sine-Gordon model is invariant under $SU(2)_q$ 
quantum group symmetry. To make this symmetry manifest, one has to 
redefine the ZF operators
\begin{equation}
\hat Z_a(\beta )=\exp \bigg( a{\beta\over 2\xi}\bigg)Z_a(\beta )\ ,
\label{sgzfredef}
\end{equation}
which implies the following redefinition of the S-matrix
\begin{equation}
\hat S_{ab}^{cd}(\beta )=
\exp \bigg( (a-c){\beta\over 2\xi}\bigg)S_{ab}^{cd}(\beta ).
\label{sgsmatredef}
\end{equation}
It can be easily proved that if we identify the vectors $\hat Z_a(\beta )
\vert{\rm vac}\rangle$ as the fundamental representation of the 
quantum group $SU(2)_q$, then the redefined S-matrix and the scattering 
relations (\ref{scattrel}) are invariant under the action of the 
quantum group. However, care must be taken since $\hat S$ is not a 
unitary matrix and the representation of the out states differs from 
that of the in states by a change $q\rightarrow 1/q$. Since 
$\vert q\vert =1$, this is equivalent to a complex conjugation. Using 
this invariance and the fact that the decomposition of the corresponding 
representations is given by ${1\over 2}\otimes{1\over 2}=0\oplus 1$ 
(we label the representations using the conventional isospin notation), 
we can extract the amplitudes in the two possible channels which turn 
out to be
\begin{equation}
S_{isospin-1}(\beta )=S_0(\beta ),\ 
S_{isospin-0}(\beta )=
{\sinh\left({\displaystyle i\pi 
-\beta\over\displaystyle\xi}\right)\over 
\sinh\left({\displaystyle i\pi 
+\beta\over\displaystyle\xi}\right)}S_0(\beta )\ ,
\label{sgchannelsmat}
\end{equation}
and $\hat S$ can be decomposed as 
\begin{equation}
\hat S=S_{isospin-0}P_0+S_{isospin-1}P_1\ ,
\label{sgprojsmat}
\end{equation}
where $P_0$ and $P_1$ are projectors on the spin-0 and spin-1 
subspace of the product of two spin-1/2 representations.

The sausage model was introduced in \cite{saus1} as a one-parameter 
family of deformations of the O(3) nonlinear sigma model.  The 
spectrum of the model consists of a massive (quantum group) isotriplet 
of scalars, capital letters $I,J,K,L$ will be used to label the internal 
quantum number which takes the values $+,0,-$. The S-matrix is invariant 
under a $U(1)$ symmetry and the charge conjugation which takes $\bar I=
-I$. The S-matrix has the following symmetries:
\begin{eqnarray}
{\cal S}_{IJ}^{KL}(\beta )={\cal S}_{\bar I\bar J}^{\bar K\bar L}(\beta )=
{\cal S}_{JI}^{LK} (\beta )={\cal S}_{KL}^{IJ}(\beta )\quad ({\rm C,P,T}) 
\ ,\nonumber \\
{\cal S}_{IJ}^{KL}(\beta )={\cal S}_{J\bar I}^{L\bar K}(i\pi-\beta )
\quad ({\rm crossing})\ .
\label{saussmatsym}
\end{eqnarray}
The independent S-matrix-elements are
\begin{eqnarray}
&&{\cal S}_{++}^{++}(\beta )={\sinh\lambda (\beta -i\pi )
\over\sinh\lambda (\beta +i\pi )} \ ,\nonumber\\
&&{\cal S}_{+0}^{0+}(\beta )=-i{\sin 2\pi\lambda\over
\sinh\lambda (\beta -2i\pi )}{\cal S}^{++}_{++}(\beta ) \ ,\nonumber \\
&&{\cal S}_{+-}^{-+}(\beta )=-{\sin\pi\lambda\sin 2\pi\lambda\over
\sinh\lambda (\beta +i\pi )\sinh\lambda (\beta -2i\pi )}\ ,\nonumber\\
&&{\cal S}_{+0}^{+0}(\beta )={\sinh \lambda\beta\over
\sinh\lambda (\beta -2i\pi )}{\cal S}^{++}_{++}(\beta ) \ ,\nonumber \\
&&{\cal S}_{00}^{00}(\beta )={\cal S}_{+0}^{+0}(\beta )+
{\cal S}_{+-}^{-+}(\beta )\ ,
\label{saussmat}
\end{eqnarray}
where $\lambda$ is a parameter of the model. This is a consistent 
S-matrix if $0\leq\lambda\leq1/2$; otherwise, there are poles 
inside the physical sheet $0\leq{\rm Im}\beta <\pi$. In \cite{saus1}
the one-loop renormalized action can be found. For our considerations 
we will need only the S-matrix (\ref{saussmat}).

The S-matrix of the sausage-model is $SU(2)_q$ invariant with 
$q=\exp [i\pi(\lambda +1)]$. When we set $\lambda\rightarrow 0$, we find 
the S-matrix of the O(3) nonlinear sigma model. One can introduce 
the ZF operators ${\cal Z}_I(\beta )$ for the sausage model similarly 
to the case of sine-Gordon theory. The basis needed to exhibit the 
quantum group symmetry is given by
\begin{equation}
\hat{\cal Z}_I(\beta )=\exp (\lambda I\beta){\cal Z}_I(\beta ).
\label{sauszfredef}
\end{equation}
The states created by the operators $\hat{\cal Z}$ form a triplet 
(isospin-1) representation of $SU(2)_q$. Similarly to the case of 
sine-Gordon model (see (\ref{sgprojsmat})), we can decompose the 
S-matrix into three channels corresponding to $1\otimes 1=0\oplus 
1\oplus 2$. The amplitudes for the three channels are
\begin{eqnarray}
&&{\cal S}_0(\beta )={\sinh\lambda (\beta +2i\pi )
\over\sinh\lambda (\beta -2i\pi )}\ ,\nonumber\\
&&{\cal S}_1(\beta )={\sinh\lambda (\beta +2i\pi )
\over\sinh\lambda (\beta -2i\pi )}
{\displaystyle\strut \sinh\lambda (\beta -i\pi )\over
\displaystyle\strut \sinh\lambda (\beta +i\pi )}\ ,
\nonumber\\
&&{\cal S}_2(\beta )={\sinh\lambda (\beta -i\pi )
\over\sinh\lambda (\beta +i\pi )}\ .
\end{eqnarray}
\section{Form-factor axioms and the free field representation 
for their solution}\label{sect3}

In Smirnov's approach to integrable field theories, the local operators 
are defined in terms of their form-factors. Form-factors are the matrix 
elements of the local operators between the vacuum and many-particle 
states:
\begin{equation}
F^{\cal O}_{i_1,...,i_n}(\beta_1,...,\beta_n)=\langle vac\vert{\cal O}(0)
\vert\beta_1,...,\beta_n\rangle_{i_1,...i_n}\ .
\label{ffactdef}
\end{equation}
By virtue of crossing symmetry, these matrix elements determine the general 
matrix element between two many-particle states and, assuming asynptotic 
completeness of Hilbert space, this means the complete characterization of 
the local operator ${\cal O}(x)$. The form-factors of a local operator must 
satisfy the following set of axioms:

1. The functions $F$ are analytic in the rapidity differences
$\beta_{ij}=
\beta_i-\beta_j$ inside the strip $0<{\rm Im}\beta <2\pi$ except for
simple poles.
The physical matrix elements are the values of the $F$'s at
real rapidity arguments ordered as $\beta_1<\beta_2<\dots <\beta_n$.

2. The form-factor of a local operator of Lorentz spin $s$ satisfies
\begin{equation}
F_{i_1,\dots ,i_n}(\beta_1+\Lambda ,\dots ,\beta_n+\Lambda )=
\exp (s\Lambda )F_{i_1,\dots ,i_n}(\beta_1 ,\dots ,\beta_n ).
\end{equation}
This means that the
form-factor of a spin $s=0$ operator depends only on
the rapidity differences.

3. Form-factors satisfy Watson's symmetry property:
\begin{eqnarray}
&&F_{i_1,\dots ,i_{j},i_{j+1},\dots ,i_n}
(\beta_1,\dots ,\beta_j,\beta_{j+1},
\dots ,\beta_n)=\nonumber \\ && S_{i_{j}i_{j+1}}^{k_{j+1}k_{j}}
F_{i_1,\dots ,k_{j+1},k_{j},\dots ,i_n}
(\beta_1,\dots ,\beta_{j+1},\beta_{j},
\dots ,\beta_n).
\end{eqnarray}

4. Form-factors satisfy the cyclic property
\begin{equation}
F_{i_1,\dots ,i_n}(\beta_1,\dots ,\beta_n+2\pi i)=
F_{i_n,i_1,\dots ,i_{n-1}}(\beta_n,\beta_1,\dots ,\beta_{n-1}).
\end{equation}
In the case of an operator which has a mutual locality index 
$\omega ({\cal O},\Psi )$ with respect to the elementary field 
$\Psi$, this axiom is modified to
\begin{equation}
F_{i_1,\dots ,i_n}(\beta_1,\dots ,\beta_n+2\pi i)=
\exp [2\pi i\omega ({\cal O},\Psi )]
F_{i_n,i_1,\dots ,i_{n-1}}(\beta_n,\beta_1,\dots ,\beta_{n-1}).
\end{equation}

5. Form-factors have kinematical singularities which are simple poles
at points where two of the rapidity arguments are displaced by $i\pi$.
The residue is given by the following formula:
\begin{eqnarray}
&&  {\rm res}_{\beta_i=\beta_n+i\pi}
F_{i_1,\dots ,i_n}(\beta_1,\dots ,\beta_n)=
-iC_{i_n,i_j^,}
F_{i_1^{,},\dots ,\hat i_j^,\dots ,i_{n-1}^{,}}(\beta_1,\dots ,
\hat\beta_j,\dots ,\beta_n)\times\nonumber \\
&&  [\delta_{i_1}^{i_1^{,}}\dots\delta_{i_{j-1}}^{i_{j-1}^{,}}
S_{i_{n-1}k_1}^{i_{n-1}^{,}i_j^{,}}(\beta_{n-1}-\beta_j)
S_{i_{n-2}k_2}^{i_{n-2}^{,}k_1}(\beta_{n-2}-\beta_j)\dots
S_{i_{j+1}i_j}^{i_{j+1}^{,}k_{n-j-2}}(\beta_{j+1}-\beta_j)\nonumber \\
&&  -S_{k_1,i_1}^{i_{j}^{,}i_1^{,}}(\beta_j-\beta_1)\dots
S_{k_{j-2}i_{j-2}}^{k_{j-3}i_{j-2}^{,}}(\beta_j-\beta_{j-2})
S_{i_ji_{j-1}}^{k_{j-2}i_{j-1}^{,}}(\beta_j-\beta_{j-1})
\delta_{i_{j+1}}^{i_{j+1}^{,}}
\dots\delta_{i_{n-1}}^{i_{n-1}^{,}} ]
\ .\label{formfactoraxioms}\end{eqnarray}
We would like to remark here that the cyclicity axiom closely 
resembles the KMS condition for finite-temperature Green functions 
of quantum field theory. We will see that the expressions for the 
solutions of the form-factor equations really look like expectation 
values in a heat bath and later we will use this analogy.

The locality theorem proved in \cite{smirnov4} ensures that the 
operators defined by solutions of the above equations are local 
(up to the mutual locality indices, if there are such). The 
importance of the form-factors is given by the fact that the 
correlation function of two local operators can be expanded 
into an infinite series in terms of their form-factors, obtained 
by inserting a complete set of many-particle states into the 
correlator.
 
In the paper of Lukyanov \cite{freefield}, a free field 
representation for the Zamolodchikov-Faddeev algebra was proposed 
to solve the above axioms. That work considered the SU(2) Thirring 
and the sine-Gordon models. Here we recall some essential facts 
about this free field realization. We do not give a full treatment 
and restrict ourselves to the formulas, used in describing sine-Gordon 
form-factors, which are  necessary in the subsequent 
sections. The interested reader is referred to \cite{freefield} 
\cite{jostfunc} for more details.

Let us introduce a field $\phi(\beta )$ with the following properties:
\begin{eqnarray}
&& [\phi(\beta_1),\phi(\beta_2)]=\ln S_0(\beta_2-\beta_1), \nonumber \\
&& \langle 0\vert\phi(\beta_1)\phi(\beta_2)\vert 0\rangle =
-\ln g(\beta_2-\beta_1),
\label{phirelations}\end{eqnarray}
where $S_0$ is the function defined in eqn. (\ref{s0def}) and we give
the formula for the function $g$ below (\ref{gfunct}).
The consistency of the two relations above requires
\begin{equation}
S_0(\beta )={g(-\beta )\over g(\beta )}.
\end{equation}

The field $\phi$ is represented (after a proper ultraviolet cut-off 
procedure) on a Fock space. We will build up operators which represent
the ZF algebra on that space.

The field
\begin{equation}
\bar\phi(\beta )=
\phi (\beta+i{\pi\over 2})+\phi (\beta-i{\pi\over 2}),
\label{phibar}\end{equation}
will be important to define the 
screening charge below. 
Its two-point functions can be written down as
\begin{eqnarray}
&&  \langle 0\vert\bar\phi(\beta_1)\phi(\beta_2)\vert 0\rangle =
 \ln w(\beta_2-\beta_1), \nonumber \\
&&  \langle 0\vert\bar\phi(\beta_1)\bar\phi(\beta_2)\vert 0\rangle =
 -\ln \bar g(\beta_2-\beta_1),
\end{eqnarray}
where we defined the following functions
\begin{eqnarray}
&&g(\beta )=({ {\displaystyle\strut \kappa}\over
 {\displaystyle\strut \Gamma} ({1\over\xi})})^{1\over 2}
{ {\displaystyle\strut \Gamma}({1\over\xi}+{i\beta\over\pi\xi})\over
{\displaystyle\strut \Gamma}({i\beta\over\pi\xi})}
\prod\limits_{l=1}^{\infty}{\displaystyle\strut  [R_l(i\pi )R_l(0)]^{1\over 2}\over 
\displaystyle\strut  R_l(\beta )} \ ,\nonumber \\
&&w(\beta )= \kappa^{-1}{\Gamma(-{1\over 2\xi}+{i\beta\over\pi\xi})
\over \Gamma(1+{1\over 2\xi}+{i\beta\over\pi\xi}) }\ ,
\nonumber \\
&&{\bar g}(\beta )= \kappa^2{i\beta\over\pi\xi}
{\Gamma(1+{1\over\xi}+{i\beta\over\pi\xi})
\over \Gamma(-{1\over\xi}+{i\beta\over\pi\xi}) }\ .
\label{gfunct}\end{eqnarray}
where $k$ is a normalization constant and $R_l$ is the function
\begin{equation}
R_l(\beta)={\Gamma({2l\over\xi}+{i\beta\over\pi\xi})
\Gamma(1+{2l\over\xi}+{i\beta\over\pi\xi})\over  
\Gamma({2l+1\over\xi}+{i\beta\over\pi\xi})
\Gamma(1+{2l-1\over\xi}+{i\beta\over\pi\xi}) }\ .
\end{equation}
Now we define Lukyanov's vertex operators: 
\begin{eqnarray}
V(\beta )&  =&\exp (i\phi (\beta ))=
(g(0))^{1\over 2}:\exp (i\phi (\beta )):, \nonumber \\
\bar V(\beta )&  =&\exp (-i\bar\phi (\beta ))=
(\bar g(0))^{1\over 2}
:\exp (-i\bar\phi (\beta )):,
\label{vertexdef}\end{eqnarray}
where : denotes an appropriate normal ordering \cite{freefield}.
These operators need to be regularized because the functions $g$
and $\bar g$ have simple zeros at $\beta =0$.  Using the analogy with 
conformal field theory we define the regularized values of the functions
$g,\bar g$ at $\beta =0$ to be
\begin{eqnarray}
&  g_{reg}(0)=&\lim\limits_{\beta\rightarrow 0}{ g(\beta )\over
 \beta}=:\rho^2 ,\nonumber \\
&  \bar g_{reg}(0)=&\lim\limits_{\beta\rightarrow 0}
{ \bar g(\beta )
\over \beta}=:\bar \rho^2.
\end{eqnarray}

The vertex operators defined in (\ref{vertexdef}) obey the
following fundamental relations:
\begin{eqnarray}
&  V(\beta_1)V(\beta_2)=&\rho^2g(\beta_2-\beta_1)
:V(\beta_1)V(\beta_2): \ ,\nonumber \\
&  \bar V(\beta_1)V(\beta_2)=&\rho\bar\rho w(\beta_2-\beta_1)
:V(\beta_1)V(\beta_2): \ ,\nonumber \\
&  \bar V(\beta_1)\bar V(\beta_2)=&\bar\rho^2
\bar g(\beta_2-\beta_1):\bar V(\beta_1)\bar V(\beta_2):\ .
\label{vertexrel}\end{eqnarray}
Eqn. (\ref{vertexrel}) plays a key role in proving the ZF
scattering relations.

Then in analogy with the Coulomb gas representation of rational 
conformal field theory one introduces a screening charge $\chi$
by the definition
\begin{equation}
\langle u\vert\chi\vert v\rangle = \eta^{-1} \langle u\vert{1\over 2\pi}
\int_C d\gamma\bar V(\gamma )\vert v\rangle ,
\end{equation}
where C is a contour specified in the following manner: assuming
that all matrix elements $\langle u\vert\bar V(\gamma )\vert v\rangle$
are meromorphic functions decreasing at the infinity faster than
$\gamma^{-1}$, the contour goes from $\Re \gamma =-\infty$ to
$\Re \gamma =+\infty$ and lies above all singularities whose positions 
depend on $\vert u\rangle$ and below all singularities depending on
$\vert v\rangle$. ($\eta$ is a normalization parameter which is chosen 
to get the correct residue for the pole on the operator product below 
(see \cite{freefield})).  

Having defined these objects, one can write down the following
operators:
\begin{eqnarray}
&  Z_{+}(\beta )=& \exp (-{\beta\over 2\xi})V(\beta ), \nonumber \\
&  Z_{-}(\beta )=& \exp ({\beta\over 2\xi})
(q^{1\over 2}\chi V(\beta )-q^{-{1\over 2}}V(\beta )\chi ).
\end{eqnarray}
Then the following propositions hold \cite{freefield}:
\hfill\break

(i) These operators satisfy the SU(2) Thirring model ZF relations.

(ii) The singular part of the operator product 
$Z_i(\beta_2)Z_j(\beta_1)$ considered as a function of the complex 
variable $\beta_2$ for real $\beta_1$ in the upper half plane
$\Im \beta_2\geq 0$ contains only one simple pole at 
$\beta_2=\beta_1+i\pi$ with residue $-iC_{ij}$.
\hfill\break\hfill\break
We introduce the following structures, specified later in details:

1. The operator of Lorentz boosts $K$ satisfying
\begin{equation}
Z_i(\beta+\Lambda )=\exp (-\Lambda K)Z_i(\beta)\exp (\Lambda K).
\end{equation}

2. A map ${\cal O}\rightarrow L({\cal O})$
from the space of local operators to the algebra
of endomorphisms of $\pi_Z$ satisfying the following two conditions
\begin{eqnarray}
  L({\cal O})Z_i(\beta )& =& Z_i(\beta )L({\cal O} ), \nonumber \\
 \exp ( \Lambda K)L({\cal O})\exp (-\Lambda K)& =
&\exp (\Lambda s({\cal O}))L({\cal O} ).
\end{eqnarray}
where $s({\cal O})$ denotes the Lorentz spin of the operator $\cal O$.

If one has these objects, then, as shown by Lukyanov \cite{freefield},
the functions
\begin{equation}
F_{i_1,\dots i_n}(\beta_1 ,\dots ,\beta_n)=
Tr_{\pi_Z} [\exp (2\pi iK)L({\cal O})
Z_{i_n}(\beta_n)\dots Z_{i_1}(\beta_1)],
\end{equation}
are solutions of the form-factor axioms.

We now have to give the definition of the dual operators
which give us the map $L$ representing the local operators 
of the model. We introduce the scalar field $\phi^\prime (\alpha )$
satisfying the same relations as $\phi$ together with its associated 
field $\bar\phi^\prime (\alpha )$.  (We give a closer definition of 
these fields shortly when we introduce an ultraviolet
regularization for the free field system.) With the help
of these new fields we define the dual fields as follows:

\begin{eqnarray}
&&\Lambda_-(\alpha )={U^\prime}(\alpha ), \nonumber \\
&&\Lambda_0(\alpha )=-{i\over\sqrt{2\cos({\pi\over\xi+1})}}
[q^\prime{\chi^\prime}{U^\prime}(\alpha )
-q^{\prime -1}{U^\prime}(\alpha ){\chi^\prime} ], \nonumber \\
&&\Lambda_+(\alpha )={1\over {2\cos({\pi\over\xi+1})}}
[q^\prime {\chi^\prime}^2{U^\prime}
(\alpha )-(q^\prime+ q^{\prime -1}){\chi^\prime}
{U^\prime}(\alpha ){\chi^\prime}+
q^{\prime -1}{U^\prime}(\alpha ){\chi^\prime}^2].
\end{eqnarray}
The dual vertex operators ${U^\prime}$ and the $\bar V^\prime$ required for
${\chi^\prime}$, satisfy similar relations as the unprimed ones, but with
the function $\bar g$ changed to
\begin{equation}
\bar g^\prime(\alpha )= \kappa^{\prime 2}{i\alpha\over\pi(\xi +1)}
{\Gamma(1-{1\over(\xi +1)}+{i\alpha\over\pi(\xi +1)})
\over \Gamma({1\over(\xi +1)}+{i\alpha\over\pi(\xi +1)}) } \ .
\end{equation}
The parameter $q^\prime$ equals 
$\exp \left[ -i\pi{\displaystyle 
\xi\over\displaystyle\xi +1}\right]$.

One can also calculate all the
normal ordering relations involving the
primed and unprimed operators merely by using the relation
\begin{equation}
\bar V^\prime(\delta )V(\beta )=\rho\bar\rho^\prime u(\beta -\delta)
:\bar V^\prime(\delta )V(\beta ):,
\end{equation}
where the function $u$ is given by
\begin{equation}
u(\beta )=k{i\beta\over 2\pi},
\end{equation}
with $k$ a constant.

Now one writes down the functions
\begin{eqnarray}
{\cal F}^{i_1,\dots i_k}_{j_1,\dots j_l}
(\alpha_1,\dots \alpha_k\vert
\beta_1,\dots \beta_n) =
Tr_{\pi_Z} [\exp (2\pi iK)\Lambda_{i_k}(\alpha_k)
\dots\Lambda_{i_1}(\alpha_1){Z}_{j_l}(\beta_l)
\dots{Z}_{j_1}(\beta_1)],
\label{formfsol}\end{eqnarray}
then these functions have to be expanded as
\begin{eqnarray}
{\cal F}^{i_1,\dots i_k}_{j_1,\dots j_l}
(\alpha_1,\dots \alpha_k\vert
\beta_1,\dots \beta_n)
= \sum\limits_{\{s_j\}} F^{\prime i_1,\dots i_k}
(\alpha_1,\dots \alpha_k\vert\{s_j\})F_{j_1,\dots j_l}(\{s_j\}\vert
\beta_1,\dots \beta_n)\times
\exp (s_1\alpha_1)\dots\exp (s_k\alpha_k),
\end{eqnarray}
where the functions $F$ are the required solutions of the form-factor
axioms. This means that the functions
$\cal F$ play the role of generating
functions for the form-factors of local operators. They describe 
$n$-particle form-factors of infinitely many operators which may 
be considered as descendents of the first one in the sense of 
\cite{smirnov5}.

The actual calculation of the matrix elements and the traces proceeds
through the procedure described in \cite{freefield}.
The free field construction needs a regularization;
we choose an ultraviolet
cut-off by taking the rapidity interval finite
\begin{equation}
-{\pi\over\epsilon}<\beta <{\pi\over\epsilon}.
\end{equation}
Then we introduce the mode expansion of the free field $\phi$ as:
\begin{equation}
\phi_\epsilon(\beta )=
{\sqrt{s}}(Q-i\beta P)+\sum\limits_{k\neq 0}
{a_k\over i\sinh (\pi k\epsilon )}\exp (ik\epsilon\beta),
\label{modexp}\end{equation}
with
\begin{equation}
s={\xi +1\over 2\xi}.
\end{equation}
The oscillator modes satisfy the commutation relation
\begin{equation}
[a_k,a_l]={\sinh{\pi k\epsilon\over 2}\sinh\pi k\epsilon\over k}
{\sinh{\pi k\epsilon\over 2}(\xi +1)\over
\sinh{\pi k\epsilon\over 2}\xi}\delta_{k,-l} \ ,
\end{equation}
while for the zero modes we have the usual canonical ones:
\begin{equation}
[P,Q]=-i.
\end{equation}
The two-point function and the
commutation relations for this field can
be calculated and it can be seen that in the
limit $i\rightarrow 0$
we recover the relations satisfyed by the field
$\phi$ (see eq. (\ref{phirelations})). The dual field is given by
\begin{equation}
\phi^\prime_\epsilon(\alpha )=-{\sqrt{s^\prime}}(Q-i\alpha P)+
\sum\limits_{k\neq 0}
{a^\prime_k\over i\sinh (\pi k\epsilon )}\exp (ik\epsilon\alpha),\ 
s^\prime ={\xi \over 2(\xi +1)},
\end{equation}
with the oscillators $a_k^\prime$ defined by the relation
\begin{equation}
a_k^\prime\sinh{\pi k\epsilon\over 2}(\xi +1)=
a_k\sinh{\pi k\epsilon\over 2}\xi \ .
\end{equation}
We introduce the conventional notation used in conformal field theory
\begin{equation}
\alpha_+=-1/\alpha_-=\sqrt{\xi +1\over\xi}\ .
\end{equation}
The dual set of operators can actually be obtained by the substitution 
$\xi\rightarrow -1-\xi$, which is known to have an analogy in conformal 
field theory (see Lukyanov \cite{freefield} for details).
The space where the operators and their duals act is defined to be
\begin{equation}
\pi^\epsilon_Z = \bigoplus\limits_{l,l^\prime\in \rm Z} 
F_{{\alpha_+l+\alpha_-l^\prime\over\sqrt{2}}} \ ,
\end{equation}
where $F_p$ denotes the Fock space
built up with the help of the creation
operators (as usual, the oscillator modes with
negativ indices) from the
ground state $\vert p\rangle$ which satisfies:
\begin{equation}
P\vert p\rangle=p\vert p\rangle.
\end{equation}
The operator $K$, of the Lorentz boost is given by
\begin{equation}
K_\epsilon=i\epsilon H +{{\alpha_++\alpha_-}\over\sqrt{2}}P,
\end{equation}
where
\begin{equation}
H={P^2-p_0^2\over2}+\sum\limits_{k=1}^\infty
{k^2\over\sinh{\pi k\epsilon\over 2}
\sinh\pi k\epsilon}a^\prime_{-k}a_k.
\end{equation}
Formally we can write
$\pi_Z=\lim\limits_{\epsilon\rightarrow 0}\pi^\epsilon_Z$.
This should be understood in the following sense:
we make all calculations
with the regularized operators and in the end take the limit
$\epsilon\rightarrow 0$.

\section{Sausage model from bootstrap fusion of sine-Gordon model}
\label{sect4}

In this section we describe how to obtain the sausage model S-matrix
introduced in section \ref{sect2} from that of the sine-Gordon theory. 
First we introduce some graphical notations. We denote the S-matrix 
by a cross:

\vskip 0.2truecm

\begin{center}
\setlength{\unitlength}{0.012500in}%
\begingroup\makeatletter
\def\x#1#2#3#4#5#6#7\relax{\def\x{#1#2#3#4#5#6}}%
\expandafter\x\fmtname xxxxxx\relax \def\y{splain}%
\ifx\x\y   
\gdef\SetFigFont#1#2#3{%
  \ifnum #1<17\tiny\else \ifnum #1<20\small\else
  \ifnum #1<24\normalsize\else \ifnum #1<29\large\else
  \ifnum #1<34\Large\else \ifnum #1<41\LARGE\else
     \huge\fi\fi\fi\fi\fi\fi
  \csname #3\endcsname}%
\else
\gdef\SetFigFont#1#2#3{\begingroup
  \count@#1\relax \ifnum 25<\count@\count@25\fi
  \def\x{\endgroup\@setsize\SetFigFont{#2pt}}%
  \expandafter\x
    \csname \romannumeral\the\count@ pt\expandafter\endcsname
    \csname @\romannumeral\the\count@ pt\endcsname
  \csname #3\endcsname}%
\fi
\endgroup
\begin{picture}(200,72)(65,735)
\thinlines
\put(245,740){\vector(-1, 1){ 60}}
\put( 75,775){\makebox(0,0)[lb]{\smash{\SetFigFont{9}{10.8}{it}$i_1$}}}
\put( 85,775){\makebox(0,0)[lb]{\smash{\SetFigFont{9}{10.8}{it}$i_2$}}}
\put( 65,765){\makebox(0,0)[lb]{\smash{\SetFigFont{14}{16.8}{it}S}}}
\put( 85,760){\makebox(0,0)[lb]{\smash{\SetFigFont{9}{10.8}{it}$i_2^\prime$}}}
\put( 75,760){\makebox(0,0)[lb]{\smash{\SetFigFont{9}{10.8}{it}$i_1^\prime$}}}
\put( 95,765){\makebox(0,0)[lb]{\smash{\SetFigFont{14}{16.8}{it}($\beta_{12}$)=}}}
\put(185,740){\vector( 1, 1){ 60}}
\put(175,800){\makebox(0,0)[lb]{\smash{\SetFigFont{9}{10.8}{it}$i_2^\prime$}}}
\put(240,755){\makebox(0,0)[lb]{\smash{\SetFigFont{9}{10.8}{it}$\beta_2$}}}
\put(175,740){\makebox(0,0)[lb]{\smash{\SetFigFont{9}{10.8}{it}$i_1$}}}
\put(250,800){\makebox(0,0)[lb]{\smash{\SetFigFont{9}{10.8}{it}$i_1^\prime$}}}
\put(185,755){\makebox(0,0)[lb]{\smash{\SetFigFont{9}{10.8}{it}$\beta_1$}}}
\put(250,740){\makebox(0,0)[lb]{\smash{\SetFigFont{9}{10.8}{it}$i_2$}}}
\end{picture}
\end{center}

This should be understood as follows: the lines are vector spaces carrying 
the internal indices of the particles. The S-matrix acts in the tensor 
product of two such spaces and carries an argument which is the rapidity 
difference of the two scattering particle. The rapidities are associated to 
the lines of the diagram. Each segment carries an index and repeated indices 
must be summed over. The order of the matrix products is represented by the 
arrows. As an example, the Yang-Baxter equation in these notations can be 
written as (omitting the rapidities of the lines, which causes no confusion)
\vskip 0.2truecm

\begin{center}
\setlength{\unitlength}{0.012500in}%
\begingroup\makeatletter
\def\x#1#2#3#4#5#6#7\relax{\def\x{#1#2#3#4#5#6}}%
\expandafter\x\fmtname xxxxxx\relax \def\y{splain}%
\ifx\x\y   
\gdef\SetFigFont#1#2#3{%
  \ifnum #1<17\tiny\else \ifnum #1<20\small\else
  \ifnum #1<24\normalsize\else \ifnum #1<29\large\else
  \ifnum #1<34\Large\else \ifnum #1<41\LARGE\else
     \huge\fi\fi\fi\fi\fi\fi
  \csname #3\endcsname}%
\else
\gdef\SetFigFont#1#2#3{\begingroup
  \count@#1\relax \ifnum 25<\count@\count@25\fi
  \def\x{\endgroup\@setsize\SetFigFont{#2pt}}%
  \expandafter\x
    \csname \romannumeral\the\count@ pt\expandafter\endcsname
    \csname @\romannumeral\the\count@ pt\endcsname
  \csname #3\endcsname}%
\fi
\endgroup
\begin{picture}(270,127)(75,665)
\thinlines
\put(180,680){\vector(-1, 1){100}}
\put(110,680){\vector( 0, 1){100}}
\put(240,680){\vector( 1, 1){100}}
\put(340,680){\vector(-1, 1){100}}
\put(310,680){\vector( 0, 1){100}}
\put(205,725){\makebox(0,0)[lb]{\smash{\SetFigFont{14}{16.8}{rm}=}}}
\put( 75,665){\makebox(0,0)[lb]{\smash{\SetFigFont{10}{12.0}{rm}$i_1$}}}
\put(105,665){\makebox(0,0)[lb]{\smash{\SetFigFont{10}{12.0}{rm}$i_2$}}}
\put(175,665){\makebox(0,0)[lb]{\smash{\SetFigFont{10}{12.0}{rm}$i_3$}}}
\put( 75,785){\makebox(0,0)[lb]{\smash{\SetFigFont{10}{12.0}{rm}$i_3^\prime$}}}
\put(110,785){\makebox(0,0)[lb]{\smash{\SetFigFont{10}{12.0}{rm}$i_2^\prime$}}}
\put( 80,680){\vector( 1, 1){100}}
\put(180,785){\makebox(0,0)[lb]{\smash{\SetFigFont{10}{12.0}{rm}$i_1^\prime$}}} 
\put(320,725){\makebox(0,0)[lb]{\smash{\SetFigFont{10}{12.0}{rm}$i_2^{\prime\prime}$}}} 
\put( 95,725){\makebox(0,0)[lb]{\smash{\SetFigFont{10}{12.0}{rm}$i_2^{\prime\prime}$}}}
\put(125,705){\makebox(0,0)[lb]{\smash{\SetFigFont{10}{12.0}{rm}$i_1^{\prime\prime}$}}}
\put(125,745){\makebox(0,0)[lb]{\smash{\SetFigFont{10}{12.0}{rm}$i_3^{\prime\prime}$}}}
\put(235,665){\makebox(0,0)[lb]{\smash{\SetFigFont{10}{12.0}{rm}$i_1$}}}
\put(310,665){\makebox(0,0)[lb]{\smash{\SetFigFont{10}{12.0}{rm}$i_2$}}}
\put(345,665){\makebox(0,0)[lb]{\smash{\SetFigFont{10}{12.0}{rm}$i_3$}}}
\put(305,785){\makebox(0,0)[lb]{\smash{\SetFigFont{10}{12.0}{rm}$i_2^\prime$}}}
\put(235,785){\makebox(0,0)[lb]{\smash{\SetFigFont{10}{12.0}{rm}$i_3^\prime$}}}
\put(335,785){\makebox(0,0)[lb]{\smash{\SetFigFont{10}{12.0}{rm}$i_1^\prime$}}}
\put(295,745){\makebox(0,0)[lb]{\smash{\SetFigFont{10}{12.0}{rm}$i_1^{\prime\prime}$}}}
\put(295,705){\makebox(0,0)[lb]{\smash{\SetFigFont{10}{12.0}{rm}$i_3^{\prime\prime}$}}}
\end{picture}
\end{center}
which means
\begin{equation}
S_{i_2^\prime i_3^\prime}
^{i_2^{\prime\prime} i_3^{\prime\prime}}(\beta_{23})
S_{i_1^{\prime} i_3^{\prime\prime}}
^{i_1^{\prime\prime} i_3}(\beta_{13})
S_{i_1^{\prime\prime} i_2^{\prime\prime}}
^{i_1 i_2}(\beta_{12})=
S_{i_1^\prime i_2^\prime}
^{i_1^{\prime\prime} i_2^{\prime\prime}}(\beta_{12})
S_{i_1^{\prime\prime} i_3^\prime}
^{i_1 i_3^{\prime\prime}}(\beta_{13})
S_{i_2^{\prime\prime} i_3^{\prime\prime}}
^{i_2 i_3}(\beta_{23}).
\label{ybeqn}\end{equation}
The bootstrap fusion means the following formal operation on the level of the 
ZF operators
\begin{equation}
{\cal Z}_I(\beta)=C^{i_1i_2}_IZ_{i_1}(\beta_1+{i\pi\over 2})Z_{i_2}(\beta_2
-{i\pi\over 2}),
\label{zffusion}\end{equation}
where $C^{i_1i_2}_I$ denotes the appropriate Clebsh-Gordan coefficients for 
the internal indices. In our case they are just the coefficients for the 
spin-$1$ representation in the product of two spin-$1/2$ representations. 
The nonzero elements are:
\begin{equation}
C^{++}_+=C^{--}_-=1\ ,\ C^{+-}_0={1\over\sqrt{q+q^{-1}}}q^{-1/2}\ , \
C^{-+}_0={1\over\sqrt{q+q^{-1}}}q^{1/2}\ .
\label{cgcoeff}\end{equation}
Eqn. (\ref{zffusion}) is formal since the rapidity difference 
just corresponds to the position of the annihilation poles 
and so there are potential divergences. 
Fortunately, in the case we investigate these singularities drop out and we 
will be able to define the ZF operators corresponding to the RHS of the 
equation \ref{zffusion} in the free field representation.
We will denote this operation graphically by
\vskip 0.2truecm

\begin{center}
\setlength{\unitlength}{0.012500in}%
\begingroup\makeatletter
\def\x#1#2#3#4#5#6#7\relax{\def\x{#1#2#3#4#5#6}}%
\expandafter\x\fmtname xxxxxx\relax \def\y{splain}%
\ifx\x\y   
\gdef\SetFigFont#1#2#3{%
  \ifnum #1<17\tiny\else \ifnum #1<20\small\else
  \ifnum #1<24\normalsize\else \ifnum #1<29\large\else
  \ifnum #1<34\Large\else \ifnum #1<41\LARGE\else
     \huge\fi\fi\fi\fi\fi\fi
  \csname #3\endcsname}%
\else
\gdef\SetFigFont#1#2#3{\begingroup
  \count@#1\relax \ifnum 25<\count@\count@25\fi
  \def\x{\endgroup\@setsize\SetFigFont{#2pt}}%
  \expandafter\x
    \csname \romannumeral\the\count@ pt\expandafter\endcsname
    \csname @\romannumeral\the\count@ pt\endcsname
  \csname #3\endcsname}%
\fi
\endgroup
\begin{picture}(175,23)(110,725)
\thinlines
\put(120,730){\line( 1, 0){100}}
\multiput(215,745)(0.40000,-0.40000){26}{\makebox(0.4444,0.6667){\SetFigFont{7}{8.4}{rm}.}}
\multiput(225,735)(-0.40000,-0.40000){26}{\makebox(0.4444,0.6667){\SetFigFont{7}{8.4}{rm}.}}
\put(120,740){\line( 1, 0){100}}
\put(225,735){\vector( 1, 0){ 55}}
\put(110,725){\makebox(0,0)[lb]{\smash{\SetFigFont{12}{14.4}{it}$i_2$}}}
\put(285,735){\makebox(0,0)[lb]{\smash{\SetFigFont{14}{16.8}{it}I}}}
\put(110,740){\makebox(0,0)[lb]{\smash{\SetFigFont{12}{14.4}{it}$i_1$}}}
\end{picture}
\end{center}
Then we can write down the graphical representation for the S-matrix 
obtained from the bootstrap fusion:
\vskip 0.2truecm

\begin{center}
\setlength{\unitlength}{0.012500in}%
\begingroup\makeatletter
\def\x#1#2#3#4#5#6#7\relax{\def\x{#1#2#3#4#5#6}}%
\expandafter\x\fmtname xxxxxx\relax \def\y{splain}%
\ifx\x\y   
\gdef\SetFigFont#1#2#3{%
  \ifnum #1<17\tiny\else \ifnum #1<20\small\else
  \ifnum #1<24\normalsize\else \ifnum #1<29\large\else
  \ifnum #1<34\Large\else \ifnum #1<41\LARGE\else
     \huge\fi\fi\fi\fi\fi\fi
  \csname #3\endcsname}%
\else
\gdef\SetFigFont#1#2#3{\begingroup
  \count@#1\relax \ifnum 25<\count@\count@25\fi
  \def\x{\endgroup\@setsize\SetFigFont{#2pt}}%
  \expandafter\x
    \csname \romannumeral\the\count@ pt\expandafter\endcsname
    \csname @\romannumeral\the\count@ pt\endcsname
  \csname #3\endcsname}%
\fi
\endgroup
\begin{picture}(165,149)(50,665)
\thinlines
\put( 90,700){\line( 1, 1){ 70}}
\put( 80,770){\line( 1,-1){ 70}}
\put(150,700){\line(-1, 1){ 70}}
\put( 90,780){\line( 1,-1){ 70}}
\put( 80,765){\line( 0, 1){ 15}}
\put( 80,780){\line( 1, 0){ 15}}
\put(145,780){\line( 1, 0){ 15}}
\put(160,780){\line( 0,-1){ 15}}
\put( 60,800){\line( 1,-1){ 20}}
\put(160,780){\line( 1, 1){ 20}}
\put( 60,680){\line( 1, 1){ 20}}
\put(160,700){\line( 1,-1){ 20}}
\put( 75,710){\line( 1, 0){ 15}}
\put( 90,710){\line( 0,-1){ 15}}
\put(150,695){\line( 0, 1){ 15}}
\put(150,710){\line( 1, 0){ 15}}
\put( 80,710){\line( 1, 1){ 70}}
\multiput( 80,700)(0.40000,0.40000){26}{\makebox(0.4444,0.6667){\SetFigFont{7}{8.4}{rm}.}}
\put(195,735){\makebox(0,0)[lb]{\smash{\SetFigFont{12}{14.4}{it}
$={\cal S}^{I_1I_2}_ {I_1^\prime I_2^\prime}(\beta_{12})$}}}
\multiput(160,700)(-0.40000,0.40000){26}{\makebox(0.4444,0.6667){\SetFigFont{7}{8.4}{rm}.}}
\put( 50,665){\makebox(0,0)[lb]{\smash{\SetFigFont{14}{16.8}{it}$I_1$}}}
\put(185,665){\makebox(0,0)[lb]{\smash{\SetFigFont{14}{16.8}{it}$I_2$}}}
\put( 45,805){\makebox(0,0)[lb]{\smash{\SetFigFont{14}{16.8}{it}$I_2^\prime$}}}
\put(185,805){\makebox(0,0)[lb]{\smash{\SetFigFont{14}{16.8}{it}$I_1^\prime$}}}
\put(105,700){\makebox(0,0)[lb]{\smash{\SetFigFont{12}{14.4}{it}$i_2$}}}
\put(125,700){\makebox(0,0)[lb]{\smash{\SetFigFont{12}{14.4}{it}$i_3$}}}
\put(155,720){\makebox(0,0)[lb]{\smash{\SetFigFont{12}{14.4}{it}$i_4$}}}
\put( 75,720){\makebox(0,0)[lb]{\smash{\SetFigFont{12}{14.4}{it}$i_1$}}}
\put( 75,750){\makebox(0,0)[lb]{\smash{\SetFigFont{12}{14.4}{it}$i_3^\prime$}}}
\put(105,770){\makebox(0,0)[lb]{\smash{\SetFigFont{12}{14.4}{it}$i_4^\prime$}}}
\put(125,770){\makebox(0,0)[lb]{\smash{\SetFigFont{12}{14.4}{it}$i_1^\prime$}}}
\put(155,750){\makebox(0,0)[lb]{\smash{\SetFigFont{12}{14.4}{it}$i_2^\prime$}}}
\end{picture}
\end{center}
This diagram should be read as follows: two incoming particle in the 
triplet representation is converted (using the fusion formula 
\ref{zffusion}) into two pairs of doublets and then scattered on each other 
(this means four crossings ie. four S-matrix factors). After that, the 
triplet part is projected out using again the fusion formula. However, care 
must be taken, since in the case of the incoming particles we make an 
embedding of the triplet representation into the product of the two 
doublet representations, while in the case of the outgoing particles we 
go the other way via projection. These two operations are adjoint to each 
other and differ not only by transposition of matrices, but (and here comes 
an essential difference from the O(3) sigma model case treated in \cite{prev}) 
also differ by a substitution $q\rightarrow 1/q$, which is here just a 
complex conjugation since $\vert q\vert =1$. We can write the fusion equation 
in the following form (using the redefinitions \ref{sgzfredef}, 
\ref{sgsmatredef} and \ref{sauszfredef})

\begin{equation}
{\hat{\cal S}}_{I_1^\prime I_2^\prime}^{I_1I_2}(\beta_{12})=
C_{I_1^\prime}^{i_1^\prime i_2^\prime}
C_{I_2^\prime}^{i_3^\prime i_4^\prime}
{\hat S}^{i_1^{\prime\prime}i_4^{\prime\prime}}_{i_1^\prime i_4^\prime}
(\beta_{12}+i\pi )
{\hat S}^{i_1^{\prime\prime}i_3^{\prime}}_{i_1 i_3^{\prime\prime}}
(\beta_{12})
{\hat S}_{i_2^{\prime}i_4^{\prime\prime}}^{i_2^{\prime\prime} i_4}
(\beta_{12})
{\hat S}_{i_2^{\prime\prime}i_3^{\prime\prime}}^{i_2 i_3}
(\beta_{12}-i\pi )
\tilde C^{I_1}_{i_1 i_2}
\tilde C^{I_2}_{i_3 i_4}
\end{equation} 
The coefficients $C$ and $\tilde C$ are given in (\ref{cgcoeff}).
We remark that it is necessary to use the redefined S-matrices because to 
apply the quantum group Clebsh-Gordan coefficients we have to go to the 
one-particle basis which is at the same time the properly normalized 
basis of the quantum group representation.

The proof of the fusion relation between the sine-Gordon and sausage model 
S-matrices proceeds as follows. The sine-Gordon S-matrix acts in the 
tensor product of two spaces; we can expand it in terms of tensor products 
of Pauli matrices acting in one space as follows:
\begin{eqnarray}
S_0(\beta ){1\over 2\sinh ({\beta-i\pi\over\xi})}\Big[ &&I_1\otimes I_2
\Big(\sinh \Big({\beta-i\pi\over\xi}\Big)
-\sinh \Big({\beta\over\xi}\Big)\Big)+\sigma_1^3\otimes
\sigma_2^3\Big(\sinh\Big({\beta-i\pi\over\xi}\Big)
+\sinh \Big({\beta\over\xi}\Big)\Big)  \nonumber\\
&&-\sinh\Big({i\pi\over\xi}\Big)
(\sigma_1^1\otimes \sigma_2^1+\sigma_1^2\otimes \sigma_2^2)\Big] .
\end{eqnarray}
The redefinition means changing the Pauli matrices by rapidity-dependent 
factors. The product of the four S-matrices can be written in the tensor 
product of four two-dimensional spaces, each S-matrix acting only in two of 
them: eg. the first S-matrix in the spaces 1 and 4, the second one in the 
spaces 1 and 3 and so on. Then one can embed the Clebsh-Gordan coefficients 
into this 16 dimensional space and carry out the matrix multiplication. The 
special structure reflected in the graphical presentation of the formula 
greatly simplifies the required computation. The relation between the 
parameters of the two models turns out to be $\lambda =1/\xi$.

Now we turn to the fusion of the ZF operators. We want to give a meaning 
to the formula \ref{zffusion}. We interpret it as the limit
\begin{equation}
{\hat{\cal Z}}_I(\beta)=\lim\limits_{\delta\rightarrow 0}
C^{i_1i_2}_I{\hat Z}_{i_1}(\beta_1+{i\pi\over 2}+\delta)
{\hat Z}_{i_2}(\beta_2 -{i\pi\over 2}-\delta).
\label{zffuslim}\end{equation}
Remarkably, all the singularities cancel and the result is a very simple 
formula for in terms of the free field representation:
\begin{eqnarray}
&&{\cal Z}_+(\beta )= \exp\Big(-{\beta\over\xi}\Big){U}(\beta ), \nonumber \\
&&{\cal Z}_0(\beta )={i\over\sqrt{2\cos({\pi\over\xi})}}
[q{\chi}{U}(\beta )
-q^{-1}{U}(\beta ){\chi} ], \nonumber \\
&&{\cal Z}_-(\beta )=-{1\over \cos({\pi\over\xi})}
\exp\Big( {\beta\over\xi}\Big) [q{\chi}^2{U}
(\beta )-(q+q^{-1}){\chi} {U}(\beta ){\chi}+q^{-1}{U}(\beta ){\chi}^2].
\end{eqnarray}

The proof of this result can be found in Appendix \ref{appa}, where we 
also outline the direct calculation of the scattering relations. They prove 
to be identical to the sausage model ZF algebra:
\begin{equation}
{\cal Z}_I(\beta_1){\cal Z}_J(\beta_2)= 
{\cal S}_{IJ}^{KL}(\beta_{12}){\cal Z}_L(\beta_2)
{\cal Z}_K(\beta_1).
\end{equation}
However, just as in the case of the O(3) sigma model \cite{prev}, the 
operator product does not have the right kinematical poles. 
Instead of having them at rapidity difference $i\pi$, it turns 
out that the singularities are situated at $2i\pi$ (see Appendix 
\ref{appc}). Nevertheless, 
if we write down the trace formula \ref{formfsol}, it will solve the cyclicity 
(by the definition of the trace) and Watson's axiom. The analytic structure 
will be corrected by a procedure similar to that in \cite{prev}.
 
\section{Evaluation of the expressions for the form-factors}
\label{sect5}

We have to calculate the traces required for obtainig the integrand of
the representation for the form-factors.
Using the coherent state representation for the traces, we find the result:
\begin{eqnarray}
&& Tr_{\pi_Z}[\exp (2\pi iK)U^\prime (\alpha_k)\dots U^\prime (\alpha_1)
\bar V^\prime (\delta_p)\dots\bar V^\prime (\delta_1)
U (\beta_n)\dots U (\beta_1)\bar V(\gamma_r)\dots\bar V(\gamma_1)]
=\nonumber \\
&&{\cal C}_1^{-n} {\cal C}_2^{n+r\over 2}
{\cal C}_2^{\prime {p+k\over 2}} 2^{-r-n}
i^{n+p+k}\eta^r\eta^{\prime -p-k}
\delta_{nr}\delta_{pk}\exp\left(
{1\over\xi}\left[\sum\limits_{j=1}^n\beta_j-
\sum\limits_{j=1}^r\gamma_j\right]+
{1\over\xi +1}\left[\sum\limits_{j=1}^p\delta_j-
\sum\limits_{j=1}^k\alpha_j\right]\right)\times\nonumber \\
&&\prod\limits_{i<j}\bar G(\beta_i-\beta_j)
\prod\limits_{i<j}\bar G(\gamma_i-\gamma_j)
\prod\limits_{i,j}\bar G^{-1}(\gamma_i-\beta_j)
\prod\limits_{i<j}\bar G^\prime(\alpha_i-\alpha_j)
\prod\limits_{i<j}\bar G^\prime(\delta_i-\delta_j)
\prod\limits_{i,j}\bar G^{\prime -1}(\delta_i-\alpha_j)\times\nonumber\\
&&\prod\limits_{i,j}\bar H(\beta_i-\alpha_j)
\prod\limits_{i,j}\bar H^{-1}(\beta_i-\delta_j)
\prod\limits_{i,j}\bar H(\gamma_i-\delta_j)
\prod\limits_{i,j}\bar H^{-1}(\gamma_i-\alpha_j).
\label{traces}\end{eqnarray}

Here we defined the new functions as:
\begin{eqnarray}
&&\bar G(\beta )= -{ {\cal C}_2\over  4}
\xi \sinh({\beta+i\pi\over\xi} )\sinh\beta, \nonumber \\
&&\bar G^\prime (\alpha )= -{\cal C}_2^\prime
{ \sinh\alpha\over (\xi +1)\sinh({\alpha +i\pi\over \xi +1})}, 
\nonumber \\
&&\bar H(\alpha )= -{ 2\over \cosh\alpha},
\label{Gbar}
\end{eqnarray}
and the constants are given by:
\begin{eqnarray}
&& {\cal C}_1=\exp\Bigg[ -\int\limits_0^\infty { dt\over t}
{\displaystyle \sinh^2 {t\over 2}\exp (-t)
\over \sinh 2t \cosh t}
{\sinh t(\xi -1)\over\sinh t\xi}\Bigg]\ ,\nonumber \\
&& {\cal C}_2=\exp\Bigg[ 4\int\limits_0^\infty { dt\over t}                    
{\displaystyle \sinh^2 {t\over 2}\exp (-t)
\over \sinh 2t \cosh t}
{\sinh t(\xi -1)\over\sinh t\xi}\Bigg]\ ,\nonumber \\
&& {\cal C}_2^\prime=
\exp\Bigg[ -4\int\limits_0^\infty { dt\over t}                    
{\displaystyle \sinh^2 {t\over 2}\exp (-t)
\over \sinh 2t \cosh t}
{\sinh t\xi \over\sinh t(\xi +1)}\Bigg]\ .
\end{eqnarray}
The trace over the zero modes contains an
infinite constant which we have set equal to unity.

We also have to specify the integration
over the variables of the screening
charges. We face the following problem:
when we take the integration contour
as given in the sine-Gordon model (see in \cite{freefield})
and try to take the limit which gives
the sausage model, new double poles arise
in the integration over the
$\gamma_i$ variables whose positions depend
on the $\beta_j$ variables.
This phenomenon was already noticed by
Smirnov \cite{smirnov4} for the case of the O(3) sigma model.
The contour gets pinched
by the two poles approaching each other and the integrals diverge.
This means that the limit to the sausage model is singular.
To deal with the integral we introduce the prescription that one
should take the coefficient of the most divergent term in the integral.

We are now to give a justification of the prescription and explain 
the appearance of double poles in the "finite temperature" expressions, 
while they are not present in the vacuum expectation values, ie. the 
"zero-temperature" matrix elements.

The basic picture is as follows. Taking the trace means imposing the 
periodicity equation (or KMS condition) included in the form-factor 
axioms. What happens is that while in calculating matrix elements 
we use Green functions of the scalar field on the plane, here we 
impose the periodcity condition. The original function ${\bar g}(\beta )$ 
has poles at the rapidity difference $0$ and $i\pi$, which will have 
all their images under the KMS and the Watson's symmetry condition. 
They can be viewed as mirror charges introduced to satisfy the boundary 
conditions for ${\bar G}(\beta )$, which is the analogue of 
finite-temperature Green function, if we take ${\bar g}$ 
to be the analogue of zero temperature one. 
Let us for simplicity take the case of the O(3) model, which 
means $\xi=\infty$. We deform the trace by substituting 
$\exp(2\pi iK)$ by $\exp((2\pi+\zeta) iK)$, where $\zeta$ is a 
small positive number. This means a deformation 
of form-factors which is the same as considered in \cite{niedermaier}.
The deformed form-factor wiil obey Watson's symmetry property and 
a deformed KMS condition:
\begin{equation}
F_{i_1,\dots ,i_n}(\beta_1,\dots ,\beta_n+(2\pi+\zeta )i)=
F_{i_n,i_1,\dots ,i_{n-1}}(\beta_n,\beta_1,\dots ,\beta_{n-1}).
\end{equation}
Looking at the explicit form of the function ${\bar G}$, when $\zeta >0$,
which is (up to a normalization factor)
\begin{equation}
{\bar G}^{(\zeta )}(\beta )={(\beta +i\pi )\sinh\left( 
{\displaystyle \pi\beta\over 
\displaystyle 2\pi +\zeta}\right) \over
\Gamma \left( 1- 
{\displaystyle\pi-i\beta \over \displaystyle 2\pi+\zeta}\right)
\Gamma \left( 1- 
{\displaystyle\pi+i\beta \over \displaystyle 2\pi+\zeta}\right)}\ ,
\end{equation}
(for the calculation see Appendix \ref{appb}), we see that the 
double pole in ${\bar G}^{-1}$ (which is relevant for the integrand 
of the contour integral over $\gamma_i$) 
arises when $\zeta \rightarrow 0$. However, the 
operator product singularities are the same for finite $\zeta$, 
ie we have a pole at $\beta_{12} =2\pi i$ and all its periodic 
images, and the generating function is regular at $\beta_{12} =\pi i$.
So we take the leading singular term of it when $\zeta\rightarrow 0$.
It will automatically obey the (undeformed) KMS condition and 
Watson's property. But its behaviour at the positions of rapidity 
difference $i\pi$ changes. In Appendix \ref{appb} it is shown that 
the nonsingular behaviour of the operator product at these points 
is to be attributed to a zero in the Green function cancelling the 
pole coming from the pinch of the contour integral. But the same 
effect, which doubles the pole in the integrand, also doubles this 
zero in the factors ${\bar G}(\beta_i-\beta_j)$ and so the net result is 
a function having simple poles at $\beta_{ij}=0, 2\pi i$ and 
simple zeros at $\beta_{ij}=\pi i$. This is displayed well in 
the structure of the generating function for the O(3) current, 
calculated in \cite{prev}, which we remind here:
\begin{equation}
{2i\pi (\beta_{21}+i\pi )\over (\beta_{21}+2i\pi )\beta_{21} }
\left[{1\over \cosh(\beta_1-\alpha )}
+{1\over \cosh(\beta_2-\alpha )}\right]
\tanh^2{\beta_{21}\over 2}\ .
\end{equation} 
To deal with this problem, we saw in \cite{prev}, that we had to 
introduce a $2\pi i$ periodic, symmetric function, which killed 
the poles at $0$ and $2\pi i$ and turned the zero at $\pi i$ 
into a pole. This was the function (here written for n-particle 
case) 
\begin{equation}
\prod\limits_{1\leq i<j\leq n}\tanh^2{\beta_{ij}\over 2}\ .
\end{equation}
The end result for the generating function of the two-particle 
form-factors of the current (and its descendents) is therefore 
\begin{equation}
{\displaystyle i\pi m
\over\displaystyle 8}\epsilon^{ABC}
{\displaystyle\beta_{12}-i\pi\over\displaystyle\beta_{12}
(2i\pi-\beta_{12})}
\tanh^2{\displaystyle\beta_{12}\over\displaystyle 2}
\left( {1\over\cosh (\alpha -\beta_1)}+
{1\over\cosh (\alpha -\beta_2)}\right) .
\end{equation}
taking into account the proper normalization factors to get the 
correct value of the charges on particle states.
In the light of the argument given above, we expect this to 
be a general prescription to get the right structure of kinematical 
poles. 

Although in the above discussion we took $\xi =\infty$, it carries 
over to finite values of $\xi$. In this case additional poles are 
introduced (both in ${\bar g}$ and ${\bar G}$) 
but they are outside the strip 
since we choose $\xi >2$ in order to be in the regime where the 
sausage model has no bound state poles. Taking suitably small
$\zeta$ it is possible to draw the contours in a way which avoids 
these additional singularities. 

For the contours of integration over the variables $\delta_j$, 
no similar problem arises and we have to draw the contours just as 
for the vacuum matrix elements but they must lie within the 
strip $-i\pi <\delta_j<i\pi$.
This completes our prescription of how to write down generating 
functions for the form-factors of the sausage model.

We would like to note that in the case of  finite $\xi$, the 
correct analogue 
of the above generating function is not just a simple continuation 
to finite $\xi$ \cite{prev}. The expression
\begin{equation}
Tr_{\pi_Z} [\Lambda_-(\alpha){\cal Z}_0(\beta_2){\cal Z}_+(\beta_1)]
\end{equation}
turns out to be zero due to the Kronecker delta arising from the zero 
mode trace. The correct analogue would be to take eg.
\begin{equation}
Tr_{\pi_Z} [\Lambda_0(\alpha){\cal Z}_+(\beta_2){\cal Z}_-(\beta_1)].
\end{equation}
If one writes down this expression, one gets a complicated integral 
formula which however can be seen to yield the correct 
O(3) value in the limit 
$\xi\rightarrow 0$ (as well as all other formulas for form-factors 
written down using the general expression (\ref{traces})).

The case $\xi=\infty$ is special since in that case the 
appropriate factor is $\delta_{2n-2r+2p-2k}$, which allows much 
more nonzero solutions \cite{prev}. 
The reason for that is that the zero-mode 
trace enforces momentum conservation for the free field. While in 
the O(3) case $\alpha_+=-\alpha_-$ and the momentum cancellation 
is possible between the vertex operator and their duals, for generic 
$\xi$ this is not the case: they have to cancel separately. Since 
we have actually Two quantum groups, one with $q=\exp i\pi\alpha_+^2$ 
for the vertex operators ${\cal Z}$ and one with $q^{\prime}=
\exp i\pi\alpha_-^2$ for their duals, 
the condition which is imposed is actually 
that we should have operators under the trace which are singlets 
separately under each of them. This is very much in the spirit 
of Smirnov's work \cite{smirnov5} and resembles his notion 
of the "main set of solutions" (which are the one which have their 
counterparts when deforming away from $q=-1$). If $\xi$ is 
rational, however, some cancellations happen and the space of 
the solutions increases, suggesting an analogy with the reduction
of sine-Gordon theory at rational values of the coupling, corresponding 
to some well-known changes in the representation theory of the quantum 
group when the value of $q$ equals a root of unity \cite{redsg}. 

\section{Conclusions and further questions}\label{sect6}

In this paper we were able to show that there exists a free field 
representation for the sausage model, very similar to the one found 
in \cite{prev} for the O(3) case. We could also identify
the origin of the problem with the kinematical singularities and the 
singularity in the fusion limit:  
the first is related to the fact that the operator product
singularities of the fused ZF operators are not the correct ones, 
while the second is due to the to the additional poles introduced 
when taking the "finite-temperature" Green function instead 
of the vacuum one. Both of them is ultimately related to 
the zero of the function ${\bar g}=\beta_{21}(\beta_{21}+i\pi)$ at  
$\beta_{12}=i\pi$. 
We would like to remark that the $\tanh^2$ factors which were 
put in by hand can be considered as arising from vertex operators of 
usual conformal free scalar fields with the two-point function 
$-\log \beta_{12}$. 
Since such vertex operators will have trivial braiding, 
one can write down a tensor product representation by tensoring our 
vertex operators with the conformal free field vertex operators. Then 
the expressions for the form-factors will automatically include the 
correction factors for the kinematical poles. However, all our attempts
to find an "irreducible" (not tensor product) representation have failed 
so far.  For this one has to change the ${\bar g}$ function, while retaining 
its property
\begin{equation}
{{\bar g}(-\beta )\over {\bar g}(\beta )}=
{ \sinh \left({\displaystyle \beta -i\pi \over\displaystyle\xi}\right)
\over 
\sinh \left({\displaystyle \beta +i\pi \over\displaystyle\xi}\right)}\ ,
\end{equation}
in such a way that we get the right operator product structure. 
One has to notice that the above condition determines ${\bar g}$ only up to 
multiplication by an even function. However it does not seem possible 
to find such a function, although we have no proof that such a function 
cannot exist. We simply have been unable to construct it so far. 
The solution, which takes tensor product representation does not seem to be 
superior to the procedure of simply introducing the additional factors. 

We have obtained a procedure with which one can (in principle) 
calculate form-factors of the sausage model. The main question, however, 
is to find a general procedure with which one can associate a similar 
free field representation to any factorizable S-matrix and in this 
way find the solutions for the form-factor axioms. The other question is 
whether the solutions obtained from the free field representation 
form a complete set. If this is true, then the method may provide a major 
step towards understanding the operator content of the models.
We also want to mention in this context the relations to the work of 
Smirnov \cite{smirnov5}. It would be interesting to analyse the 
form-factors from the point of KZ equations and following the lines 
of \cite{mussardo1,mussardo2,smirnov6} to obtain a counting of 
the local fields in the 
model. It also seems worthwile to examine the rational values of $\xi$ 
and to see what sort of structure emerges in that case.

\acknowledgements

The authors would like to thank M. R. Niedermaier for useful discussion. 
This work was partially supported by the Hungarian National Science and
Research Foundation. 

\appendix

\section{Proof of the bootstrap fusion and the Zamolodchikov-Faddeev 
algebra relations}\label{appa}

In this appendix we outline the proof of the bootstrap fusion and the 
ZF relation. Starting with the bootstrap fusion, the only nontrivial 
relation to prove is:
\begin{equation}
{\hat{\cal Z}}_+(\beta)= C_{--}^{-}
{\hat Z}+_{i_1}(\beta_1+{i\pi\over 2})
{\hat Z}+_{i_2}(\beta_2 -{i\pi\over 2}).
\end{equation}
Writing out in details:
\begin{eqnarray}
&&\exp\Big({\beta\over\xi}\Big) [q{\chi}^2{U}
(\beta )-(q+q^{-1}){\chi} {U}(\beta ){\chi}+q^{-1}{U}(\beta ){\chi}^2]
=\nonumber\\
&&\exp \left({\beta_+\over 2\xi}\right)
(q^{1\over 2}\chi V(\beta_+ )-q^{-{1\over 2}}V(\beta_+ )\chi )
\exp \left({\beta_-\over 2\xi}\right)
(q^{1\over 2}\chi V(\beta_- )-q^{-{1\over 2}}V(\beta_- )\chi ),
\end{eqnarray}
where $\beta_\pm =\beta\pm i\pi /2$. 

First, we prove the equality for the integrands using the 
normal ordering procedure. We symmetrize in the arguments of the 
screening charges, since they are to be integrated out. We get 
the relation:
\begin{eqnarray}
&&(q{\bar g}^{-1}(\beta-\gamma_1 ){\bar g}^{-1}(\beta -\gamma_2 )-
(q+q^{-1}){\bar g}^{-1}(\gamma_1-\beta ){\bar g}^{-1}(\beta -\gamma_2 )
+q^{-1}{\bar g}^{-1}(\gamma_1-\beta ){\bar g}^{-1}(\gamma_2 -\beta ))
{\bar g}(\gamma_2-\gamma_1)+\nonumber\\
&&({\rm similar\  term\  with\  }\gamma_1\leftrightarrow\gamma_2)=
\nonumber\\
&&{\bar g}(\gamma_2-\gamma_1)w(\gamma_2-\beta_+)w(\beta_--\gamma_1)
(q^{1\over 2}w(\beta_+-\gamma_1)-q^{-{1\over 2}}w(\gamma_1-\beta_+))
(q^{1\over 2}w(\beta_--\gamma_2)-q^{-{1\over 2}}w(\gamma_2-\beta_-))+
\nonumber\\
&&({\rm similar\  term\  with\  }\gamma_1\leftrightarrow\gamma_2)\ ,
\label{integrand}\end{eqnarray}
which can be proven after some elementary, but lengthy calculations.
Then one proves the equality the integrals, which amounts to 
checking that the contours can be deformed into each other.  

For the proof of the ZF relations one actually 
uses only the ratio ${\bar g}(\beta )/ {\bar g}(-\beta )$. 
Hence one can substitute the functions 
${\bar g}$ by other functions having the same ratio, 
eg. $\bar G$ in the proof.  
The proof is essentially the same as presented in \cite{prev}. 
In that work, we could only work out the 
proof up to the identities containing two screening charges. Since 
then, improving our MAPLE algorithms, we have been able to prove the 
ZF relations for all combinations of the $\cal Z$ operators, 
including those with three and four screening charges, in the 
general sausage model case. The procedure is very similar 
to the proof of the fusion relation, but the formulas are even 
more lengthy and we do not attempt to describe it here. For the 
O(3) case it was discussed in some details in \cite{prev}.

\section{Evaluation of traces using coherent oscillator methods}
\label{appb}

Here we show how to calculate the traces involved in the form-factors 
using coherent oscillator methods. We will take a very simple example 
and calculate in the O(3) limit, when $\xi=\infty$. We will also 
calculate only the non-zero mode part of the trace, since it is 
the most interesting contribution when we discuss the analytical 
structure in the main text. We will take the case when the 
regularizing parameter $\zeta$ is finite and reproduce the function
${\bar G}^(\zeta )$ used in the main text.
The expression we want to calculate is:
\begin{equation}
Tr(e^{\omega iK}e^{i{\bar\phi}(\beta_1)}e^{i{\bar\phi}(\beta_2)})
\end{equation}
with $\omega=2\pi-\zeta$ introduced for convenience. We omitted 
the indices $\epsilon$ from $K$ and ${\bar\phi}$ since it is clear 
that we use their regularized version and this will cause no 
confusion.

We use the following redefinitions. First, we change our oscillator 
to have the usual normalization:
\begin{equation}
[\alpha_n, \alpha_m]=m\delta_{m+n.0}.
\end{equation}
These are related to $a_m$ in the following way:
\begin{equation}
a_m=\sqrt{N_m(\epsilon )}\alpha_m, \quad 
a_{-m}=\sqrt{N_m(\epsilon )}\alpha_{-m},
\end{equation}
where 
\begin{equation}
N_m(\epsilon )={ \sinh{\pi m\epsilon\over 2}\sinh \pi m\epsilon
\over m}e^{\pi m\epsilon\over 2}.
\end{equation}
As usual, the operators $\alpha_m$ and $\alpha_{-m}$ are Hermitian 
conjugates. The non-zero mode part of the operator $K$ looks very 
familiar:
\begin{equation}
K=i\epsilon \sum\limits_{m=1}^\infty m\alpha_{-m}\alpha_m+\ 
{\rm zero\  mode\  terms}
\end{equation}
We normal order under the trace to get:
\begin{equation}
{\bar g}_(\beta_{21})
Tr(e^{\omega iK}:e^{i{\bar\phi}(\beta_1)}e^{i{\bar\phi}(\beta_2)}:),
\end{equation}
and put in the explicit expressions (\ref{phibar}), (\ref{modexp}).
Observe that the trace factorizes into independent pieces 
over each harmonic oscillator. We evaluate it in the basis of coherent 
states. The relevant formulas are:
\begin{eqnarray}
\alpha\vert\lambda )=\lambda\vert\lambda ), \ 
(\lambda\vert\alpha^\dagger =(\lambda\vert{\bar\lambda}, \
x^{\alpha^\dagger\alpha}\vert\lambda )=\vert x\lambda ), \nonumber \\
Tr {\cal O}=\int{d\lambda d{\bar\lambda}\over\pi}
e^{-\lambda{\bar\lambda}}(\lambda\vert{\cal O}\vert\lambda ).
\end{eqnarray}
Here we omitted the mode index $m$ since the above formulas are the 
same for each oscillator mode. Performing the calculation for each 
mode separately, we get a Gaussian integral, which can be evaluated 
easily. Putting together all the contributions we find
\begin{equation}
{\bar g}(\beta_{21})
\prod\limits_{m=1}^\infty (1-e^{-\omega\epsilon m})^{-1}
\exp\left( -{2N_m(\epsilon)e^{-\omega\epsilon m}\over
\sinh^2{\pi m\epsilon\over 2}(1-e^{-\omega\epsilon m}) }\right)
\prod\limits_{m=1}^\infty 
\exp\left( -{2N_m(\epsilon)e^{-\omega\epsilon m}\cosh (im\epsilon
\beta_{12})\over
\sinh^2{\pi m\epsilon\over 2}(1-e^{-\omega\epsilon m}) }\right)
\end{equation}
Now we substitute $N_m(\epsilon )$ and then expand 
\begin{equation}
(1-e^{-\omega\epsilon m})^{-1}=\sum\limits_{n=0}^\infty
(e^{-\omega\epsilon n})^m
\end{equation}
in the exponent. We can then sum up in $m$ using the formula
\begin{equation}
\sum\limits_{m=1}^\infty {x^m\over m}=\log (1-x).
\end{equation}
Ignoring the rapidity independent parameters (which can be 
assembled into the coefficient ${\cal C}_2$), we arrive at 
an infinite product formula:
\begin{equation}
\prod_{n=0}^\infty\left[ 
(1-e^{-\alpha\epsilon n+(\pi -\alpha )\epsilon +i\epsilon m\beta_{12}})
(1-e^{-\alpha\epsilon n+(\pi -\alpha )\epsilon -i\epsilon m\beta_{12}})
(1-e^{-\alpha\epsilon n-\alpha\epsilon +i\epsilon m\beta_{12}})
(1-e^{-\alpha\epsilon n-\alpha\epsilon -i\epsilon m\beta_{12}})
\right].
\end{equation}
Using the definition of the deformed gamma function
\begin{equation}
\Gamma_\epsilon (x)=[1-e^{-2\pi\epsilon}]^{1-x}
\prod\limits_{n=0}^\infty
{1-\exp (-2\pi\epsilon (n+1))\over
1-\exp (-2\pi\epsilon (x+n))}\ ,
\end{equation}
and the fact, that for $\epsilon\rightarrow 0$ it tends to the usual
gamma function, we get for ${\bar G}^{(\zeta )}(\beta_{12})$ the expression 
(apart from constant factors):
\begin{equation}
{(\beta_{21}+i\pi)\sinh\left(
{\displaystyle\pi\beta_{21}\over\displaystyle\omega}\right)\over
\Gamma\left(\displaystyle 1-{\pi\over\omega}
+{i\beta_{21}\over\omega}\right)
\Gamma\left(\displaystyle 1-{\pi\over\omega}
+{i\beta_{21}\over\omega}\right) }\ ,
\end{equation}
which in the limit $\zeta\rightarrow 0$ gives back the $\xi =\infty$
case of (\ref{Gbar}).

\section{A sample calculation of the operator product singularities}
\label{appc}

Here we sketch the calculation of operator product singularities 
on the example of the product $U(\beta_1)\chi\chi U(\beta_2)$ in the 
O(3) symmetric case. This product is given by (up to constant factors)
\begin{eqnarray}
&&(\beta_{21}+i\pi )\beta_{21}
\int{d\gamma_1\over 2\pi}\int{d\gamma_2\over 2\pi}\Bigg[
(\gamma_{21}+i\pi )\gamma_{21}
{1\over (\gamma_1-\beta_1+i\pi)(\gamma_1-\beta_1)}
{1\over (\gamma_2-\beta_1+i\pi)(\gamma_2-\beta_1)}\times\nonumber\\
&&{1\over (\beta_2-\gamma_1+i\pi)(\beta_2-\gamma_1)}
{1\over (\beta_2-\gamma_2+i\pi)(\beta_2-\gamma_2)} \Bigg] \ ,
\end{eqnarray}
with the contours specified by the prescription given in the main
text. We have possible pinch singularities at 
$\beta_2=\beta_1$, $\beta_2=\beta_1-i\pi$ and $\beta_2=\beta_1-2i\pi$.
We can analyse their contribution by substituting 
$\beta_1=\beta_2+i\epsilon$, $\beta_1=\beta_2+i\pi+i\epsilon$ 
and $\beta_1=\beta_2+2i\pi+i\epsilon$ respectively 
into the integral and studying the 
limit $\epsilon\rightarrow 0$. One finds a regular behaviour in the 
first two cases, which is caused by the zero of the preintegral factor, 
cancelling the pinch contribution. Only the third case will give 
a singular contribution, which will correspond to a first order pole 
in the operator product. Other operator products can be analysed 
similarly.

\end{document}